\documentclass{article}
\usepackage[utf8]{inputenc}
\usepackage{amsmath}
\usepackage{amssymb}
\usepackage{braket}
\usepackage{graphicx}
\usepackage{color}
\usepackage{ulem}

\newcommand{\beq}[1]{\begin{equation}}
\newcommand{\eeq}{\end{equation}}
\newcommand{\bea}[1]{\begin{eqnarray}}
\newcommand{\eea}{\end{eqnarray}}

\title{Propelling force from asymmetrically excited quantum vacuum with conventional mirrors}
\author{Yu-Song Cao\footnote{caoyusong15@mails.ucas.ac.cn}$~^{1,2}$, YanXia Liu\footnote{yxliu-china@ynu.edu.cn}$~^{2}$, Ding-Fang Zeng\footnote{dfzeng@bjut.edu.cn}$~^{1}$}
\date{}

\begin{document}

\maketitle

\noindent $~^{1}$\small{School of Physics and Optoelectronic Engineering, Beijing University of Technology, Beijing, China}

\noindent $~^{2}$\small{School of Physics and Astronomy, Yunnan University, Kunming 650091, PR China}

\begin{abstract}
Investigations show that a time-varying $\delta-\delta'$ mirror gives rise to asymmetrical vacuum radiation on its two sides, enabling one to gain propelling force from the quantum vacuum excitation. In this work, we propose a design of Casimir device to gain propelling force by exciting vacuum asymmetrically with conventional $\delta$ mirrors. This device is experimentally feasible. It consists of a cavity made up of two mirrors, a perfectly reflective one on the left and another one with time dependent transparency on the right. All particles generated from the vacuum are ultimately right-moving, so the cavity obtains a left-pointing propelling force.
\end{abstract}

\section{Introduction}
    In 1948, Casimir effect was proposed as an observable consequence of quantum fluctuation in vacuum \cite{casimir}. About ten years later, this effect was reported to be directly observed in experiment \cite{58}. In 1969, G. T. Moore predicted that particles can be generated from the quantum vacuum inside a cavity with variable length \cite{moore}. This phenomenon is known as dynamical Casimir effect (DCE), sometimes also called vacuum radiation. Several years later, calculation suggests that DCE can also occur with just one oscillating mirror \cite{D25,A54,Neto94}.

Unfortunately, the experimental observation of DCE has proven to be much more challenging than its static sibling. Estimation shows that to obtain a detectable DCE signal, the oscillation frequency of a microwave cavity wall is required to be $\sim1$ GHz, which is far beyond the reach of current mechanical technology. Recently, it is proposed that a flying plasma mirror accelerated by intense laser may serve as a candidate to meet such extreme requirement, whose experimental and theoretical analysis were discussed in \cite{plasma,D103}. Faced with such a road block, researchers devised an alternative way to observe DCE experimentally. Instead of a moving mirror, they proposed using a spatially fixed mirror with time-varying transparency to do the trick. Because it would be much more practical to modulate the mirror's transparency with high frequency \cite{L62,G95}. Working along this line, it is noteworthy that in 2011 the DCE radiation was observed in superconducting quantum interference device (SQUID), where the mirror's transparency is simulated by the inductance of SQUID \cite{nature}. For more recent advances the readers are referred to \cite{L124,josephson}.

In DCE, the energy of radiated particles comes from the external driving source of the mirror. In return, the mirror experiences a back-reaction force against its change during the DCE emission. This back-reaction force is sometimes also called the vacuum friction \cite{D25,Neto94,0101068,0101069,rev71}. Generally, in the context of DCE, the mirrors are modelled by Dirac $\delta$ function potentials in the lagrangian formalism \cite{ann1,ann2}, which can perfectly simulate the transparence and reflection of the conventional mirrors on the macroscopic level. For a $\delta$ mirror, the transparencies and reflectivities are identical on its both sides. Consequently the emission spectra of DCE on both sides are also the same \cite{D87,D94}. This implies that during the DCE emission, the averaged back-reaction force on the mirror should be zero. While the fluctuation part of the force will cause the mirror to undergo Browinian motion around the initial position \cite{0101068,0101115,gour99}. Recently, a novel type of mirrors modelled by so called generalized Robin boundary condition attracts the researchers' interests. In mathematical form, this boundary condition serves as a $\delta-\delta'$ potential \cite{D94,D91,intA,rep,jmaa,pla,D84}. The most profound feature of the $\delta-\delta'$ mirror is the asymmetry of transparencies on its different sides \cite{D94,physics}. As a result, the emission spectra of DCE are also different on the two sides \cite{D94,D105,D102}. In \cite{D102}, the authors pointed out this asymmetric DCE spectra will lead to non-vanishing averaged back-reaction force on the mirror, giving the motion of the mirror a preferred direction. So far, discussions on $\delta-\delta'$ mirrors are purely mathematical and theoretical. Their experimental realization remains a challenge \cite{D872}.

In this work, we design a device with conventional $\delta$ mirrors, whose experimental realization is already well known, to obtain asymmetrical DCE together with directed mean back-reaction force. The lesson we learned from the $\delta-\delta'$ mirror is that, as long as the DCE spectra on the different sides of the device is asymmetrically engineered, a non-zero averaged back-reaction force would be possible \cite{D102}. Here, instead of one $\delta-\delta'$ mirror, we use two conventional $\delta$ mirrors, one with time-varying transparency and one with fixed transparency. The mirror with fixed transparency will modify the DCE spectra of the other and reflect the emitted particles, resulting in asymmetric particle flux, and more importantly, momentum flux on different sides of the cavity. By momentum conservation, the non-null mean back-reaction force will be present. For simplicity, we let the mirror on the left be perfectly reflective, while the mirror on the right have a time-varying transparency. It is found that no particles are created on the left side of the cavity and all the particles created travel rightward ultimately. Furthermore, the calculation shows the particle number and radiation energy does not grow monotonically with the driving frequency. This is different from the one-mirror models. Another feature we find is as a result of resonance transparency, the radiated particles does not contain components with resonance frequencies of the cavity. Finally, we present a unified method to compute the averaged propelling force on the mirrors in the domains of time as well as frequency.

This paper is organised as follows. In Sec.2, after a brief introduction of the model, we solve the field equation via scattering approach and do the quantization. In Sec.3, we compute the DCE spectra of particle number, energy and momentum. In Sec.4, we calculate the propelling force. Some conclusion remarks and discussion are given in Sec.5.

Throughout the whole paper, the units will be chosen so that $\hbar=c=1$.

\section{General framework of the scattering approach}

\begin{figure}
\begin{center}
\includegraphics[width=.89\textwidth]{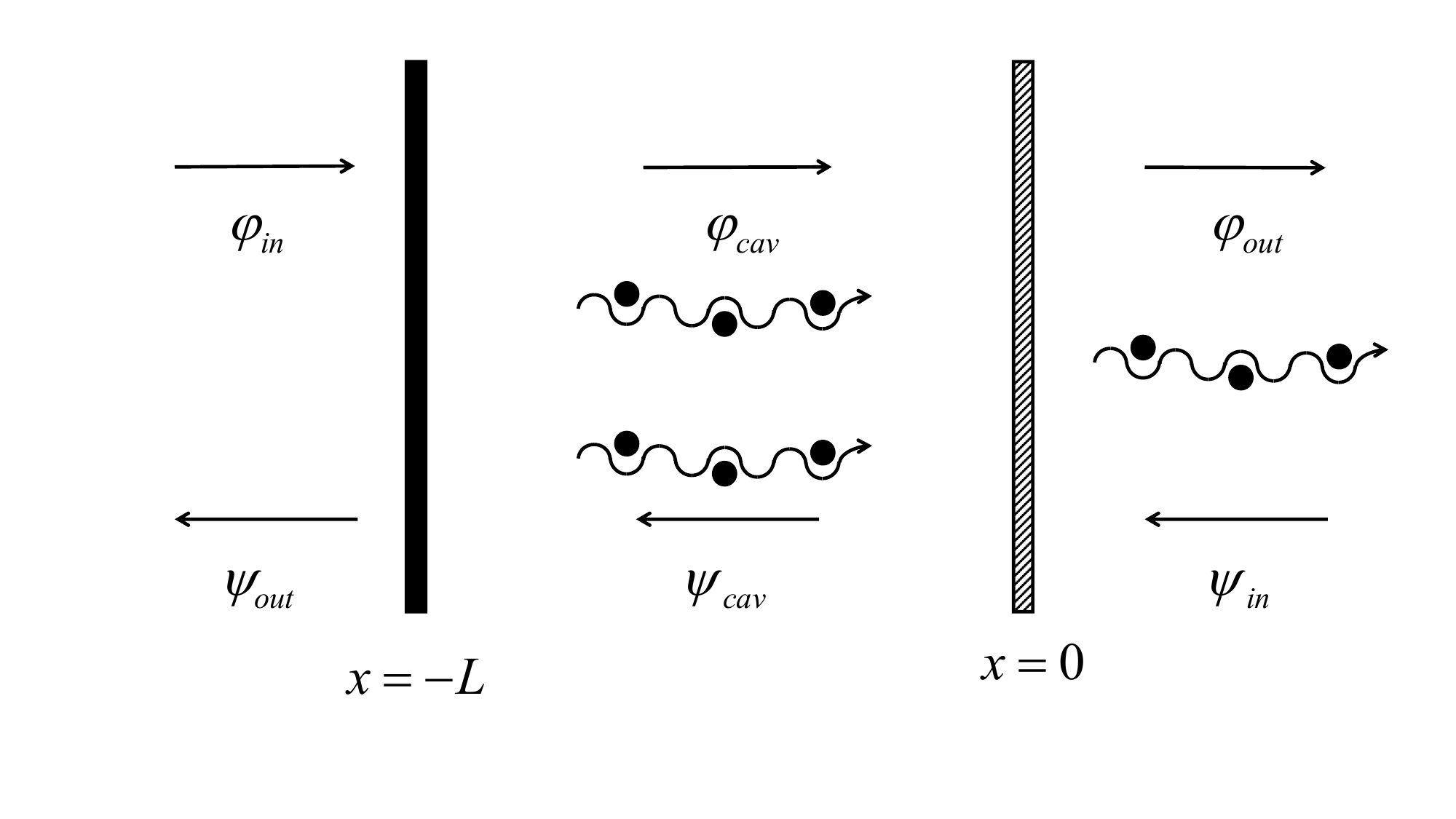}
\caption{Schematic of the system. The subscript denotes the ingoing, outgoing and cavity field, respectively. When the coupling strength of the right mirror varies, particles are created from vacuum and ultimately travels rightward, as shown by the wavy lines and the black dots.}
\label{fig:1}
\end{center}
\end{figure}

    In $(1+1)$ dimensional quantum field theory, the field can be written as the combination of left-moving and right-moving part
\bea{}
\phi(t,x)=\varphi(t-x)+\psi(t+x).
\eea{}
This can be rearranged as a column vector
\bea{}
\Phi(t,x)=\left(
\begin{array}{c}
\varphi(t-x)\\
\psi(t+x)
\end{array}
\right).
\eea{}
To characterize scattering process, the partial Fourier transformation is required, which reads
\bea{}
\label{eq:pf}
\phi(t,x)=\int\frac{d\omega}{2\pi}\phi[\omega,x]e^{-i\omega t},
\eea{}
where functions in the domains of time and frequency are distinguished by the parenthetic and square brackets. This notation will be used as default in the sequel.

The system we are interested in is a cavity interacting with a real massless scalar field. For simplicity and without loss of generality, we work in $(1+1)$ dimensions as sketched in Fig.\ref{fig:1}. In this model, both walls of the cavity are modelled by $\delta$ potentials. The Lagrangian density has the following form
\bea{}
\label{eq:lagrangian}
\mathcal{L}=\frac{1}{2}\partial_{\mu}\phi\partial^{\mu}\phi+\frac{g}{2}\delta(x+L)\phi^{2}+\frac{\lambda(t)}{2}\delta(x)\phi^{2},
\eea{}
where $g$ is the coupling strength between the left mirror and the $\phi$ field while that between the right mirror and the field is modulated by $\lambda(t)$. The field equation following this lagrangian can be written as
\bea{}
\label{eq:fieldEquation}
\square\phi(t,x)=g\delta(x)\phi(t,x)+\lambda(t)\delta(x)\phi(t,x),
\eea{}
where $\square$ is the d'Alembert operator. Solutions to this equation can be decomposed into three spatial parts
\bea{}
\label{eq:decompose}
\phi(t,x)=\theta[-(x+L)]\phi_{-}(t,x)+\theta(-x)\theta(x+L)\phi_{cav}(t,x)+\theta(x)\phi_{+}(t,x),
\eea{}
as can be seen from Fig.\ref{fig:1}.

After partial Fourier transformation \eqref{eq:pf}, Eq.\eqref{eq:decompose} gives
\begin{subequations}
\bea{}
&&\phi_{+}(t,x)=\int\frac{d\omega}{2\pi}[\varphi_{out}[\omega]e^{i\omega x}+\psi_{in}[\omega]e^{-i\omega x}]e^{-i\omega t},\\
&&\phi_{cav}(t,x)=\int\frac{d\omega}{2\pi}[\varphi_{cav}[\omega]e^{i\omega x}+\psi_{cav}[\omega]e^{-i\omega x}]e^{-i\omega t},\\
&&\phi_{-}(t,x)=\int\frac{d\omega}{2\pi}[\varphi_{in}[\omega]e^{i\omega x}+\psi_{out}[\omega]e^{-i\omega x}]e^{-i\omega t},
\eea{}
\end{subequations}
as one can infer from Fig.\ref{fig:1}. It is straightforward to see that
\begin{subequations}
\label{eq:region}
\bea{}
&&\phi_{+}[\omega,x]=\varphi_{out}[\omega]e^{i\omega x}+\psi_{in}[\omega]e^{-i\omega x},\\
&&\phi_{cav}[\omega,x]=\varphi_{cav}[\omega]e^{i\omega x}+\psi_{cav}[\omega]e^{-i\omega x},\\
&&\phi_{-}[\omega,x]=\varphi_{in}[\omega]e^{i\omega x}+\psi_{out}[\omega]e^{-i\omega x}.
\eea
\end{subequations}

The scattering on the left mirror \cite{0101067} can be written as
\bea{}
\label{eq:phil}
\Phi_{out}^{L}[\omega,-L]=S^{L}[\omega]\Phi_{in}^{L}[\omega,-L],
\eea{}
where $S^{L}[\omega]$ is the local scattering matrix, with
\bea{}
\label{eq:scat}
\Phi_{in}^{L}[\omega,-L]=\left(
\begin{array}{c}
\varphi_{in}[\omega]e^{-i\omega L}\\
\psi_{cav}[\omega]e^{i\omega L}
\end{array}
\right),\Phi_{out}^{L}[\omega,-L]=\left(
\begin{array}{c}
\varphi_{cav}[\omega]e^{-i\omega L}\\
\psi_{out}[\omega]e^{i\omega L}
\end{array}
\right).
\eea
By references \cite{D94,0101067}, the explicit form of the local scattering matrix reads
\bea{}
\label{eq:s}
S^{L}[\omega]=\left(
\begin{array}{cc}
s_{+}[\omega]&r_{+}[\omega]\\
r_{-}[\omega]&s_{-}[\omega]
\end{array}
\right),
\eea{}
where $s_{\pm}[\omega]=\frac{\omega}{\omega-ig},r_{\pm}[\omega]=\frac{ig}{\omega-ig}$ are the transparency and reflectivity. The subscripts $\pm$ denote the scattering data from the right and left, respectively. In the limit $g\to0$, we have a completely transparent mirror with $s_{\pm}[\omega]=1,r_{\pm}[\omega]=0$. While in the limit $g\to+\infty$, the mirror becomes totally reflective $s_{\pm}[\omega]=0,r_{\pm}[\omega]=-1$, where the minus sign comes from the half-wave loss. The function $\lambda(t)$ is assumed to be of the form $\lambda(t)=\lambda_{0}f(t)$, where $\lambda_{0}\ll1$ is a small parameter and $f(t)$ is an arbitrary function of time satisfying $|f(t)|\leq1$. The total reflective limit $g\to+\infty$ can be identified with the Dirichlet boundary condition. Working in this limit will reduce the computation load greatly. In the following, this limit will be taken so that
\bea{}
\label{eq:Smatrix}
\left(
\begin{array}{c}
\varphi_{cav}[\omega]\\
\psi_{out}[\omega]
\end{array}
\right)
&&=e^{i\eta\omega L}\left(
\begin{array}{cc}
&-1\\
-1&
\end{array}
\right)e^{-i\eta\omega L}\left(
\begin{array}{c}
\varphi_{in}[\omega]\\
\psi_{cav}[\omega]
\end{array}
\right)
\\
\nonumber
&&\hspace{-5mm}=\left(
\begin{array}{cc}
&-e^{2i\omega L}\\
-e^{-2i\omega L}&
\end{array}
\right)\left(
\begin{array}{c}
\varphi_{in}[\omega]\\
\psi_{cav}[\omega]
\end{array}
\right),
\eea
where $\eta=\text{diag}(1,-1)$ comes from the spatial factors $e^{\pm i\omega L}$ in Eq.\eqref{eq:scat}.

The scattering on the right mirror can be studied with the help of field equation, which in our case splits into two unrelated spatial regions $x<-L$ and $x>-L$. In $x>-L$, the field equation becomes
\bea{}
\label{eq:eom}
\square\phi(t,x)=\lambda(t)\delta(x)\phi(t,x)
\eea{}
with Dirichelet boundary condition imposed at $x=-L$. Partially Fourier transform this equation we have
\bea{}
-\partial_{x}^{2}\phi[\omega,x]+\delta(x)\int\frac{d\omega'}{2\pi}\lambda[\omega-\omega']\phi[\omega',x]=\omega^{2}\phi[\omega,x].
\eea{}
Then we integrate across $x=0$ to get
\bea{}
\label{eq:emmm}
-\partial_{x}\phi[\omega,0^{+}]+\partial_{x}\phi[\omega,0^{-}]+\int\frac{d\omega'}{2\pi}\lambda[\omega-\omega']\phi[\omega',0]=0,
\eea{}
where the continuity of the field function $\phi[\omega,0^{+}]=\phi[\omega,0^{-}]$ is used. Finally using Eqs.\eqref{eq:region} and (\ref{eq:emmm}), we will obtain
\begin{subequations}
\label{eq:mc}
\bea{}
&&\hspace{-5mm}\varphi_{cav}[\omega]+\psi_{cav}[\omega]=\varphi_{out}[\omega]+\psi_{in}[\omega],
\\
&&\hspace{-5mm}\varphi_{out}[\omega]=\psi_{in}[\omega]{+}\varphi_{cav}[\omega]{-}\psi_{cav}[\omega]
{+}\frac{\lambda_{0}}{i\omega}\int\frac{d\omega'}{2\pi}f[\omega{-}\omega'](\varphi_{cav}[\omega']{+}\psi_{cav}[\omega']).
\eea{}
\end{subequations}
Eq.\eqref{eq:mc} specifies the map between $\Phi_{in}^{R}[\omega']$ and $\Phi_{out}^{R}[\omega]$, which are defined by
\bea{}
\label{eq:phir}
&\Phi_{in}^{R}[\omega']=\left(
\begin{array}{c}
\varphi_{cav}[\omega']\\
\psi_{in}[\omega']
\end{array}
\right),\Phi_{out}^{R}[\omega]=\left(
\begin{array}{c}
\varphi_{out}[\omega]\\
\psi_{cav}[\omega]
\end{array}
\right).
\eea{}
Thus one can say that Eq.\eqref{eq:mc} contains all the information of the scattering matrix of the right mirror.

Combine Eq.\eqref{eq:mc} with Eq.\eqref{eq:Smatrix}, we have
\begin{subequations}
\bea{}
&&\varphi_{cav}[\omega]=-e^{2i\omega L}\psi_{in}[\omega]+e^{2i\omega L}\frac{\lambda_{0}}{2i\omega}\int\frac{d\omega'}{2\pi}f[\omega-\omega'](e^{-2i\omega'L}{-}1)\varphi_{cav}[\omega'],\\
&&\psi_{cav}[\omega]=\psi_{in}[\omega]+\frac{\lambda_{0}}{2i\omega}\int\frac{d\omega'}{2\pi}f[\omega-\omega'](1-e^{2i\omega'L})\psi_{cav}[\omega'].
\eea{}
\end{subequations}
This leads to the following integral equation
\bea{}
\label{eq:phiout}
\varphi_{out}[\omega]=-e^{2i\omega L}\psi_{in}[\omega]+\frac{\lambda_{0}(1-e^{2i\omega L})}{2i\omega}\int\frac{d\omega'}{2\pi}f[\omega-\omega'](\varphi_{out}[\omega']+\psi_{in}[\omega']),
\eea{}
which can be expanded in series of $\lambda_{0}$. To the second order, we have
\bea{}
&&\varphi_{out}[\omega]=-e^{2i\omega L}\psi_{in}[\omega]+\frac{\lambda_{0}(1-e^{2i\omega L})}{2i\omega} \int\frac{d\omega'}{2\pi}f[\omega-\omega'](1-e^{-2i\omega'L})\psi_{in}[\omega']
\label{eq:integral}
\\
&&-\frac{\lambda_{0}^{2}(1-e^{2i\omega L})}{2\omega}\int\frac{d\omega'}{2\pi}\frac{d\omega''}{2\pi}f[\omega-\omega']f[\omega'-\omega'']\frac{(1-e^{2i\omega' L})(1-e^{-2i\omega'' L})}{2\omega'}\psi_{in}[\omega''].
\nonumber
\eea{}

Together with Eq.(\ref{eq:Smatrix}) the global scattering matrix $S$ of the cavity can be obtained in series form
\bea{}
\Phi_{out}[\omega]=S_{0}[\omega]\Phi_{in}[\omega]+\int\frac{d\omega'}{2\pi}S_{1}[\omega,\omega']\Phi_{in}[\omega']+\int\frac{d\omega'}{2\pi}\frac{d\omega''}{2\pi}S_{2}[\omega,\omega',\omega'']\Phi_{in}[\omega''],
\eea{}
where
\begin{subequations}
\label{eq:S1}
\bea{}
&&\hspace{-5mm}S_{0}[\omega]=\left(
\begin{array}{cc}&-e^{2i\omega L}\\
-e^{-2i\omega L}&\end{array}
\right),\\
&&\hspace{-5mm}
S_{1}[\omega,\omega']=\left(\!\!
\begin{array}{cc}
0&-\frac{i\lambda_{0}(1-e^{2i\omega L})}{2\omega}f[\omega-\omega'](1-e^{-2i\omega'L})\\
0&0\end{array}
\right),\\
&&\hspace{-5mm}S_{2}[\omega,\omega',\omega'']=\left(
\begin{array}{cc}
0\!\!&-\frac{\lambda_{0}^{2}(1-e^{2i\omega L})(1-e^{2i\omega' L})}{4\omega\omega'}
f[\omega{-}\omega']f[\omega'{-}\omega''](1{-}e^{-2i\omega'' L})\\
0\!\!&0\end{array}\!\!\!
\right).
\eea{}
\end{subequations}

So far we have been working in the classical scattering theory. In order to discuss the DCE, the field must be quantized. The left and right going field will be quantized as
\begin{subequations}
\bea{}
&&\varphi[\omega]=\sqrt{\frac{1}{2|\omega|}}\left(\theta(\omega)a_{\omega}+\theta(-\omega)a_{-\omega}^{\dagger}\right),\\
&&\psi[\omega]=\sqrt{\frac{1}{2|\omega|}}\left(\theta(\omega)b_{\omega}+\theta(-\omega)b_{-\omega}^{\dagger}\right),
\eea{}
\end{subequations}
respectively. Note that operators $a$ and $b$ are for the same type of particles, the only difference between them is the direction they travel in.

\section{Vacuum radiation}

\begin{figure}
\begin{center}
\includegraphics[width=.59\textwidth]{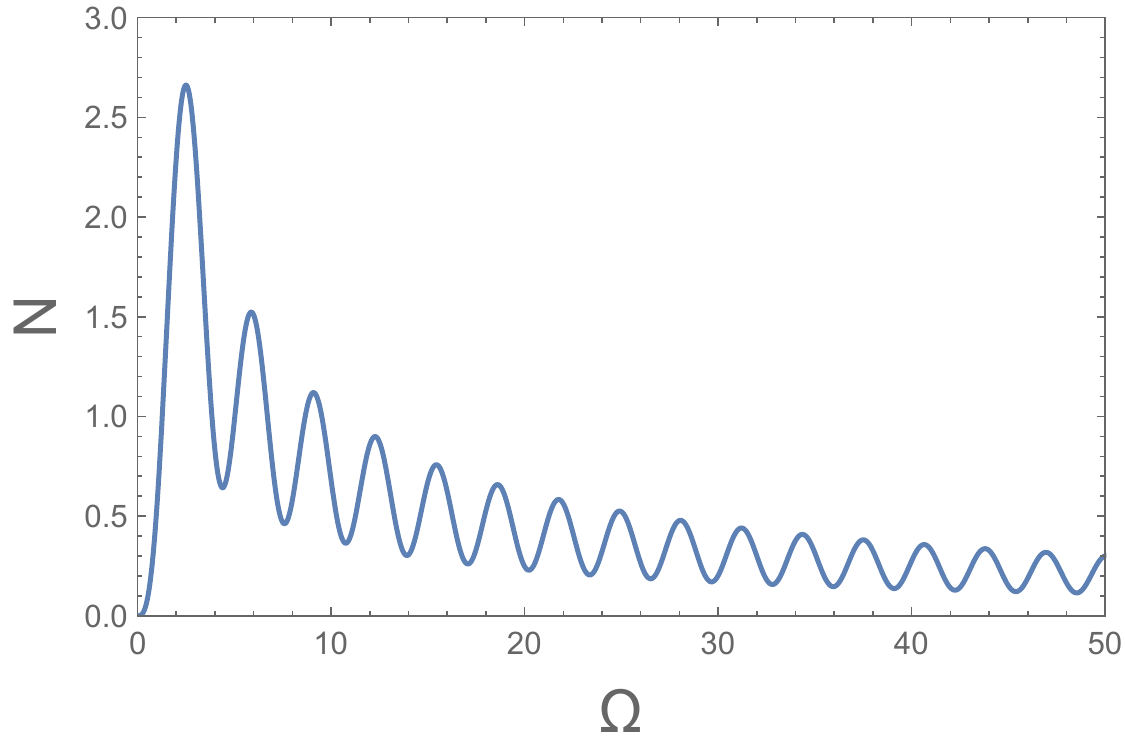}
\caption{Total number of particles with respect to driving frequency. The parameters are chosen as $\lambda_{0}=1$ and $L=1$.}
\label{fig:number}
\end{center}
\end{figure}

\begin{figure}
\begin{center}
\includegraphics[width=.59\textwidth]{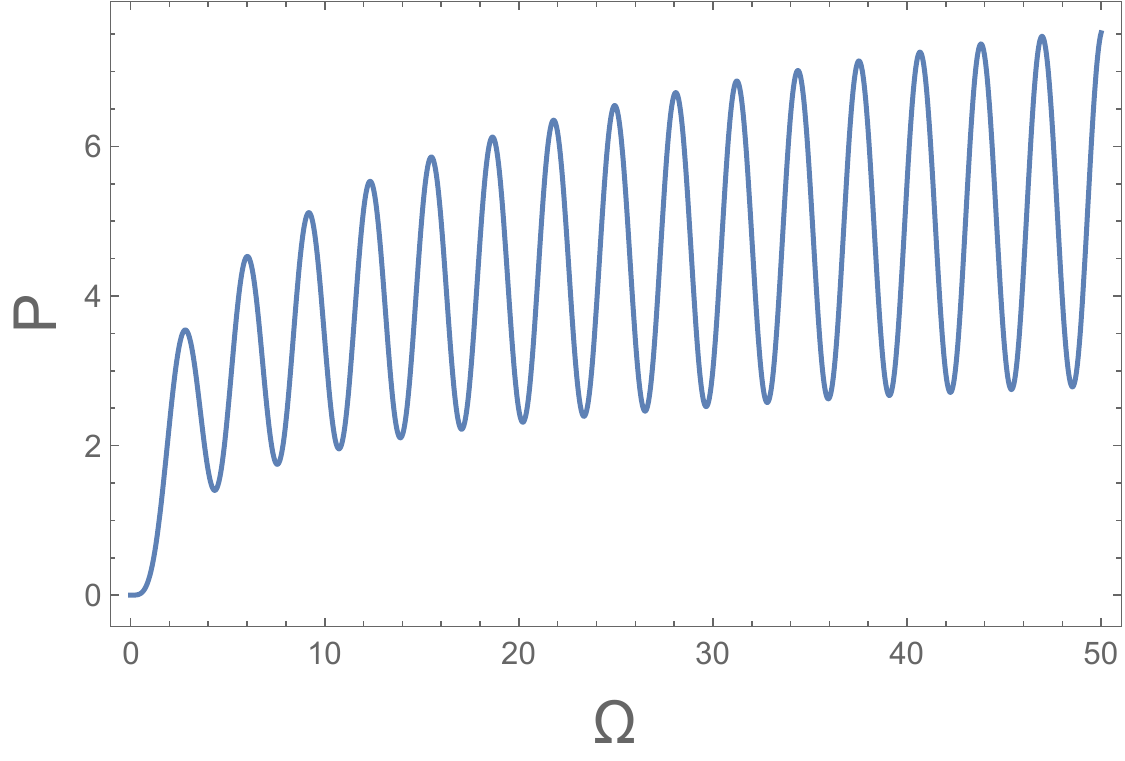}
\caption{Total momentum of particles with respect to driving frequency. The parameters are chosen as $\lambda_{0}=1$ and $L=1$.}
\label{fig:power}
\end{center}
\end{figure}

Now we ask the modulation function to satisfy $f(t)=0$ outside the time interval $[-T,T]$. In this case, before $t=-T$, we have a cavity with total transparent right mirror and a field in its vacuum state. After $t=T$, the right mirror of the cavity is also totally transparent and there will be a bunch of right moving particles in the field state, as sketched in Fig.(\ref{fig:1}). The spectra of the vacuum radiation can be calculated via \cite{L77}
\bea{}
\label{eq:spectra}
n(\omega)=2\omega\text{Tr}\bra{0_{in}}\Phi_{out}[-\omega]\Phi_{out}^{T}[\omega]\ket{0_{in}}=\int_{-\infty}^{0}d\omega'\frac{\omega}{\omega'}\text{Tr}(S[-\omega,\omega']S^{T}[\omega,-\omega']),
\eea{}
where the property
\bea{}
S[-\omega,\omega']=S^{*}[\omega,-\omega']
\eea{}
is used. Eq.(\ref{eq:Smatrix}) tells us that no particles are emitted in the left side of the cavity $x<-L$. With Eq.(\ref{eq:integral}), the emission spectra can be calculated to the second order of $\lambda_{0}$
\bea{}
\label{eq:emit}
n(\omega)=\frac{\lambda_{0}^{2}}{4\omega}|1-e^{2i\omega L}|^{2}\int_{0}^{+\infty}d\omega'\frac{|f[\omega+\omega']|^{2}|1-e^{2i\omega' L}|^{2}}{\omega'}.
\eea{}
The right hand side of Eq.(\ref{eq:emit}) vanishes when $\omega_{n}=\frac{n\pi}{L}$ and $\omega'_{n}=\frac{n\pi}{L}$, suggesting that particles of frequencies in resonance with the cavity take no part in the scattering process. This can be understood by the hole theory in which the vacuum is described as a multi-particle state, where all the negative energy levels are occupied. The DCE can thus be viewed as a scattering process of negative energy $\omega'$ particle scattered into positive energy $\omega$ state, leaving a hole in the negative energy sea. This hole, from the perspective of an observer, behaves like a particle with positive energy $-\omega'$. In this picture, it can be seen from Eq.\eqref{eq:phiout} that the modes with frequencies $\frac{n\pi}{L}$ does not participate in the interaction with the right mirror because the interaction $\lambda(t)\delta(x)\phi^{2}(x)$ is always equal to $0$ for those frequencies. With Eq.(\ref{eq:emit}), the averaged total energy and momentum of the radiated particles can thus be calculated as
\begin{subequations}
\label{eq:momentum}
\bea{}
&&E=\int_{0}^{+\infty}d\omega \omega n(\omega),\\
&&P=\int_{0}^{+\infty}d\omega \omega n(\omega).
\eea{}
\end{subequations}
The fact that $P\neq0$ gives the evidence that the averaged propelling force is not null. For illustration, we consider the damped sinusoidal modulation function
\bea{}
f(t)=e^{-\frac{|t|}{T}}\cos(\Omega t),
\eea{}
where $\Omega$ is the driving frequency and the condition
\bea{}
\lambda_{0}\Omega\gg 1
\eea{}
is required to avoid mathematical complexity \cite{A54}. In this case, the total number of particles radiated and their total momentum are given by
\begin{subequations}
\bea{}
&&N=\int_{0}^{\Omega}d\omega\frac{\lambda_{0}^{2}}{4\omega}|1-e^{2i\omega L}|^{2}\frac{|1-e^{2i(\Omega-\omega)L}|^{2}}{\Omega-\omega},\\
&&P=\int_{0}^{\Omega}d\omega\frac{\lambda_{0}^{2}}{4}|1-e^{2i\omega L}|^{2}\frac{|1-e^{2i(\Omega-\omega)L}|^{2}}{\Omega-\omega}.
\eea{}
\end{subequations}
Numerical features of these functions are plotted in Figs.(\ref{fig:number}) and (\ref{fig:power}), respectively. As we can see from Fig.(\ref{fig:number}), the particle number does not grow monotonically with the driving frequency's increasing, as one would expect in the one mirror models. There exists a critical frequency $\omega_{c}=\frac{\pi}{L}$, beyond which larger frequency tends to give less particles. As a illustration of this fact, we display in Fig.(\ref{fig:power}) the variation of the DCE particles as the driving frequency increases. From the figure we easily see that when $\omega>\omega_{c}$, increasing driving frequency is not an efficient way to acquire stronger propelling force. Both Figs.(\ref{fig:number}) and (\ref{fig:power}) exhibit the oscillation feature of period $\frac{\pi}{L}$. This can be understood with the help of Eq.(\ref{eq:emit}), where the emission spectra contains factors like $|1-e^{2i\omega L}|$ and $|1-e^{2i(\Omega-\omega)L}|$, implying resonance features of period $\frac{\pi}{L}$

\section{Propelling force}

Eq.(\ref{eq:momentum}) has demonstrated the existence of non-zero averaged propelling force. In this section, we will derive its explicit formula. To begin with, the momentum density can be written as
\bea{}
T_{01}(t,x)=T_{10}(t,x)=|\varphi'(t-x)|^{2}-|\psi'(t+x)|^{2}=P(t,x).
\eea{}
The corresponding vacuum expectation value can be obtained by
\bea{}
\langle P(t,x)\rangle=\text{Tr}[\eta\partial_{t}\partial_{t'}\langle \Phi(t,x)\Phi^{T}(t',x)\rangle]|_{t=t'}.
\eea{}
In the frequency domain, we have
\bea{}
\langle P[\omega,x]\rangle=-\int\frac{d\omega'}{2\pi}\omega'(\omega-\omega')\text{Tr}[\eta\langle\Phi[\omega',x]\Phi^{T}[\omega-\omega',x]\rangle].
\eea{}
The force on the mirrors comes from the difference of the radiation pressure on the left and right side. The averaged force can be expressed in terms of $P(t,x)$ \cite{D102}
\begin{subequations}
\bea{}
&&F(t,-L)=\langle P_{in}(t,-L)-P_{out}(t,-L)\rangle,\\
&&F(t,0)=\langle P_{in}(t,0)-P_{out}(t,0)\rangle,
\eea{}
\end{subequations}
where $F(t,-L)$ is that exerted on the left mirror and $F(t,0)$ is that on the right mirror. With Eqs.\eqref{eq:phil} and \eqref{eq:phir}, we have
\begin{subequations}
\bea{}
&&\langle P_{in}(t,-L)\rangle=\text{Tr}[\eta\partial_{t}\partial_{t'}\langle \Phi_{in}^{L}(t,-L)(\Phi_{in}^{L}(t',-L))^{T}\rangle]|_{t=t'},\\
&&\langle P_{out}(t,-L)\rangle=\text{Tr}[\eta\partial_{t}\partial_{t'}\langle \Phi_{out}^{L}(t,-L)(\Phi_{out}^{L}(t',-L))^{T}\rangle]|_{t=t'},\\
&&\langle P_{in}(t,0)\rangle=\text{Tr}[\eta\partial_{t}\partial_{t'}\langle \Phi_{in}^{R}(t,0)(\Phi_{in}^{R}(t',0))^{T}\rangle]|_{t=t'},\\
&&\langle P_{out}(t,0)\rangle=\text{Tr}[\eta\partial_{t}\partial_{t'}\langle \Phi_{out}^{R}(t,0)(\Phi_{out}^{R}(t',0))^{T}\rangle]|_{t=t'}.
\eea{}
\end{subequations}
The total propelling force on the cavity is simply
\bea{}
F(t)=F(t,-L)+F(t,0).
\eea{}
Combining with Eqs.(\ref{eq:integral}) and (\ref{eq:Smatrix}), the propelling force can be calculated order by order in $\lambda_{0}$
\begin{subequations}
\bea{}
&&F[\omega,-L]=F^{(0)}[\omega,-L]+F^{(1)}[\omega,-L]+F^{(2)}[\omega,-L]+\mathcal{O}(\lambda_{0}^{3}),\\
&&F[\omega,0]=F^{(0)}[\omega,0]+F^{(1)}[\omega,0]+F^{(2)}[\omega,0]+\mathcal{O}(\lambda_{0}^{3}).
\eea{}
\end{subequations}
Obviously, we have $F^{(0)}[\omega,-L]=F^{(0)}[\omega,0]=0$. To the first order of $\lambda_{0}$, we have
\begin{subequations}
\label{eq:f1}
\bea{}
&&\hspace{-5mm} F^{(1)}[\omega,-L]=\frac{\lambda_{0}f[\omega]}{4i}\bigg(\int_{\omega}^{+\infty}\frac{d\omega'}{2\pi}[(e^{i(2\omega'-\omega)L}-1)-e^{i\omega L}(1-e^{2i(\omega'-\omega)L})]
\\
\nonumber
&&+\int_{0}^{+\infty}\frac{d\omega'}{2\pi}[e^{i\omega L}(1-e^{-2i\omega' L})-e^{-i(\omega+2\omega')L}(1-e^{2i\omega' L})]\bigg),
\\
&&\hspace{-5mm} F^{(1)}[\omega,0]=\frac{\lambda_{0}f[\omega]}{4i}\bigg\{\int_{\omega}^{+\infty}\frac{d\omega'}{2\pi}\Big[(1{-}e^{2i(\omega'{-}\omega)L}){+}(1{-}e^{2i\omega' L})(e^{2i(\omega{-}\omega')L}{-}1)
\\
&&-e^{2i\omega' L}(1-e^{-2i(\omega'-\omega)L})\Big]+\int_{0}^{+\infty}\frac{d\omega'}{2\pi}[e^{2i(\omega-\omega')L}(1-e^{2i\omega' L})
\nonumber
\\
&&-(e^{-2i\omega' L}-1)(1-e^{2i\omega' L})-(1-e^{-2i\omega' L})]\bigg\}.
\nonumber
\eea
\end{subequations}
Expressions in Eq.(\ref{eq:f1}) are plagued by infinities from the upper bound of the integrals. The way to remedy this problem is simple. For real mirrors, they become totally transparent when the field mode frequency exceeds their plasma frequency. Thus, to obtain a physically meaningful expression for the propelling force, a cutoff should be introduced in Eq.(\ref{eq:f1}). With Eq.(\ref{eq:f1}), it's easy to prove
\bea{}
\int_{-\infty}^{+\infty}dtF^{(1)}(t)=0.
\eea{}
This is consistent with Eq.(\ref{eq:emit}) that the lowest order contribution to the DCE starts at $\lambda_{0}^{2}$. At the second order of $\lambda_{0}$,  after some lengthy algebra, the propelling force can be computed
\bea{}\label{eq:f2l}
&&\hspace{-5mm}F^{(2)}[\omega,-L]=-\frac{\lambda_{0}^{2}f[\omega]e^{i\omega L}}{4}\int\frac{d\omega'}{2\pi}\int_{0}^{+\infty}\frac{d\omega''}{2\pi}\frac{(2-e^{2i\omega'' L}-e^{-2i\omega'' L})f[\omega'-\omega'']}{\omega''}
\\
&&\hspace{-5mm}-\frac{\lambda_{0}^{2}}{8}\int_{0}^{+\infty}\frac{d\omega'}{2\pi}\int\frac{d\omega''}{2\pi}f[\omega-\omega'-\omega'']f[\omega'+\omega'']\bigg[\frac{e^{i\omega L}(1-e^{2i\omega''L})(1-e^{-2i\omega' L})}{\omega'\omega''}
\nonumber
\\
&&\hspace{-5mm}+\frac{e^{i(\omega-2\omega'+2\omega'')L}(e^{2i\omega' L}-1)(e^{-2i\omega'' L}-1))}{\omega'\omega''}\bigg]
\nonumber
\\
&&\hspace{-5mm}+\frac{\lambda_{0}^{2}}{8}\int_{0}^{\omega}\frac{d\omega'}{2\pi}\int\frac{d\omega''}{2\pi}f[\omega'-\omega'']f[\omega-\omega'+\omega'']\bigg[\frac{e^{i\omega L}(1-e^{2i\omega''L})(2-e^{2i(\omega'-\omega)L})}{\omega''(\omega'-\omega)}
\nonumber
\\
&&\hspace{-5mm}-\frac{e^{i(4\omega'+2\omega''-3\omega)L}(e^{-2i\omega'' L}-1)(e^{-2i(\omega'-\omega)L}-1)}{\omega''(\omega'-\omega)}\bigg],
\nonumber
\eea
and
\bea{}\label{eq:f2r}
&&\hspace{-5mm}F^{(2)}[\omega,0]=\frac{\lambda_{0}^{2}f[\omega]}{4}\int\frac{d\omega'}{2\pi}\int_{0}^{+\infty}\frac{d\omega''}{2\pi}f[\omega'-\omega'']\bigg(\frac{2-e^{2i\omega'' L}-e^{-2i\omega'' L}}{\omega''}
\\
&&\hspace{-5mm}-\frac{(2-e^{2i\omega' L}-e^{-2i\omega' L})(2-e^{2i\omega'' L}-e^{-2i\omega'' L})(2-e^{2i(\omega-\omega')L}-e^{-2i(\omega-\omega')L})}{\omega''}\bigg)
\nonumber
 \\
&&\hspace{-5mm}-\frac{\lambda_{0}^{2}}{8}\int_{\omega}^{+\infty}\!\!\frac{d\omega'}{2\pi}\!\int\!\frac{d\omega''}{2\pi}f[\omega'{-}\omega'']f[\omega{-}\omega'{+}\omega'']\bigg[\frac{e^{2i(\omega'{+}\omega'')L}(e^{{-}2i\omega'' L}{-}1)(e^{{-}2i(\omega'{-}\omega)L}{-}1)}{\omega''(\omega-\omega')}
 \nonumber
 \\
&&\hspace{-5mm}-\frac{e^{2i(\omega-\omega')L}(1-e^{2i\omega' L})(1-e^{2i\omega'' L})(1-e^{2i(\omega-\omega')L})+(1-e^{2i\omega'' L})(1-e^{-2i(\omega-\omega')L})}{\omega''(\omega-\omega')}\bigg]
\nonumber
\\
&&\hspace{-5mm}+\frac{\lambda_{0}^{2}}{8}\int_{0}^{+\infty}\!\!\frac{d\omega'}{2\pi}\!\int\!\frac{d\omega''}{2\pi}f[\omega{-}\omega'{-}\omega'']f[\omega'{+}\omega'']\bigg[\frac{e^{2i(\omega{-}\omega'{+}\omega'')L}(e^{2i\omega' L}{-}1)(e^{{-}2i\omega'' L}{-}1)}{\omega'\omega''}
\nonumber
\\
&&\hspace{-5mm}+\frac{e^{2i\omega' L}(1-e^{2i(\omega-\omega')L})(1-e^{2i\omega' L})(1-e^{2i\omega'' L})+(1-e^{-2i\omega' L})(1-e^{2i\omega'' L})}{\omega'\omega''}\bigg].
\nonumber
\eea

\section{Conclusion and discussion}

In this work, we designed a Casimir device made up of two conventional $\delta$ mirrors, instead of one $\delta-\delta'$ mirror, to acquire directed mean back-reaction force from asymmetric DCE. This means our scheme can be realized with current experimental platforms, for example \cite{nature,L124,josephson}. Our calculation shows that the averaged back-reaction force from DCE is non-null. Thus it will propel the device towards a preferred direction. This follows from the asymmetry of the DCE spectra on the left and right side of the cavity. The induced motion will accumulate in time due to the non-null averaged back-reaction force, which makes our device a candidate for the verification of motion induced by exciting quantum vacuum. We found that the particle number created from DCE does not grow monotonically as the driving frequency increases. Instead, it tends to decrease above a critical value $\omega_{c}=\frac{\pi}{L}$. Above this critical value, the propelling force will not significantly grow by further increasing the driving frequency.

We can define the propelling efficiency as $\eta=\frac{|P|}{W}$ to quantify how much of the energy invested to DCE is converted into propelling force, where $|P|$ is the absolute value of the total momentum of the created particles and $W$ is the total work the done by external source modulating the reflectivity of the mirror. In practical situation, only a portion of the work will be converted to the radiated particles due to the inevitable losses. We can write $E=kW$, where $E$ is the total radiated energy given by Eq.(\ref{eq:momentum}b) and $k<1$ is the ratio of the energy converted to the radiation. In our calculation, the device works at the highest efficiency $\eta_{max}=k$ according to Eq.\eqref{eq:momentum}. In the following we briefly discuss two factors that may influence this efficiency: the transparency of the left mirror and the mass of the fluctuating field.

When the transparency of the left mirror is non-zero, some of the particles created from DCE will ultimately travel leftward. So the emission spectra of the DCE will become $n(\omega)=n_{L}(\omega)+n_{R}(\omega)$, where $n_{L,R}(\omega)$ are the emission spectra on the left and right side, respectively. As a result, the efficiency will become
\begin{equation}
\eta=k\frac{\int_{0}^{+\infty}d\omega[n_{L}(\omega)-n_{R}(\omega)]}{\int_{0}^{+\infty}d\omega [n_{L}(\omega)+n_{R}(\omega)]},
\end{equation}
which is less than $\eta_{max}$ as long as $n_{L}(\omega)\neq 0$. In the limit $g\to0$ where the left mirror becomes totally transparent. In this case, the emission spectra become symmetric $n_{L}(\omega)=n_{R}(\omega)$, corresponding to zero efficiency $\eta=0$, which means the averaged back-reaction force will not propel the device at all. This is consistent with the previous results obtained in \cite{D87}.

When the mass of the vacuum fluctuating field is introduced, the efficiency of the device can be calculated by our procedure routinely
\begin{equation}
\eta=k\frac{\int_{0}^{+\infty}d\omega\sqrt{\omega^{2}-m^{2}}}{\int_{0}^{+\infty}d\omega \omega}<k.
\end{equation}
This means that if we bump equal amount of energy into the vacuum, the one uses massless field as working medium will acquire stronger propelling force than that with massive field. This is because part of the energy will become the rest mass of the radiated particles, which contributes negatively to the propelling force.

\section*{Acknowledgements}
    Y.-S. Cao thanks K. L. Wu and X. Y. Guo for helpful discussion. This research was funded by the National Natural Science Foundation of China under Grant No. $11875082$ and No. $12204406$.

\end{document}